\documentclass[reprint,aps,pra]{revtex4-2}
\usepackage{amsmath}
\allowdisplaybreaks[4]
\usepackage{multirow}
\usepackage{amsfonts}
\usepackage{xcolor}
\usepackage{comment}
\usepackage{graphicx}
\usepackage{amsmath}
\usepackage{siunitx}
\usepackage{lineno}
\usepackage{float}
\usepackage{ulem}
\usepackage[
    pdfauthor={Teng Zhang},
    pdftitle={Saturating the quantum limit with broadband signal recycling regime },
    colorlinks=true,
    linkcolor=black,
    urlcolor=blue,
    citecolor=blue
]{hyperref}

\DeclareSIUnit{\sqrthz}{\ensuremath{\sqrt{\text{\hertz}}}}

\begin{document}

\renewcommand\texteuro{FIXME}

\allowdisplaybreaks[4]
\title{A Broadband Signal Recycling Scheme for Approaching the Quantum Limit from Optical Losses}

\author{Teng Zhang}
\author{Joe Bentley}
\author{Haixing Miao}

\affiliation{$^1$School of Physics and Astronomy, and Institute of Gravitational Wave Astronomy, University of Birmingham, Edgbaston, Birmingham B15\,2TT, United Kingdom}

\begin{abstract}
Quantum noise limits the sensitivity of laser interferometric gravitational-wave detectors. Given the state-of-the-art optics, the optical losses define the lower bound of the best possible quantum-limited detector sensitivity. In this work, we come up with a broadband signal recycling scheme which gives potential solution to approaching this lower bound by converting the signal recycling cavity to be a broadband signal amplifier using an active optomechanical filter. We will show the difference and advantage of such a scheme compared with the previous white light cavity scheme using the optomechanical filter in [Phys.Rev.Lett.115.211104 (2015)].  The drawback is that the new scheme is more susceptible to the thermal noise of the mechanical oscillator.

\end{abstract}
\maketitle
\section{Introduction}

The ground-based laser interferometric gravitational-wave detectors operate at frequencies from several Hz to several kHz. The quantum noise is one of the most important noises which limits the detector sensitivity over the entire frequency band. 
To enhance the quantum-limited sensitivity, the first limit we encounter is the \textit{standard quantum limit}, which is related to the \textit{Heisenberg's uncertainty principle} of the optical quadratures \,\cite{braginsky_khalili_thorne_1992}.
It is defined as the minimal sum of the quantum shot noise and radiation pressure noise (when uncorrelated) at each frequency, given different optical power. However, it does not describe the best achievable quantum-limited sensitivity. The next limit is the \textit{quantum Cram\'er-Rao bound}~\cite{HELSTROM1967101}, or the so called energetic/fundamental quantum limit~\cite{Energetic,PhysRevLett.106.090401,PhysRevLett.119.050801}. This bound can be infinitely suppressed by improving the energy fluctuation inside the arm cavities, \textit{e.g.} 
enhancing the optical power and using quantum squeezing (anti-squeezing). It was recently realised that practical lower bound of the quantum-limited sensitivity comes from the optical-loss induced quantum dissipation\,\cite{PhysRevX.9.011053}.

\begin{figure}[b]
\centering
  \includegraphics[width=1\columnwidth]{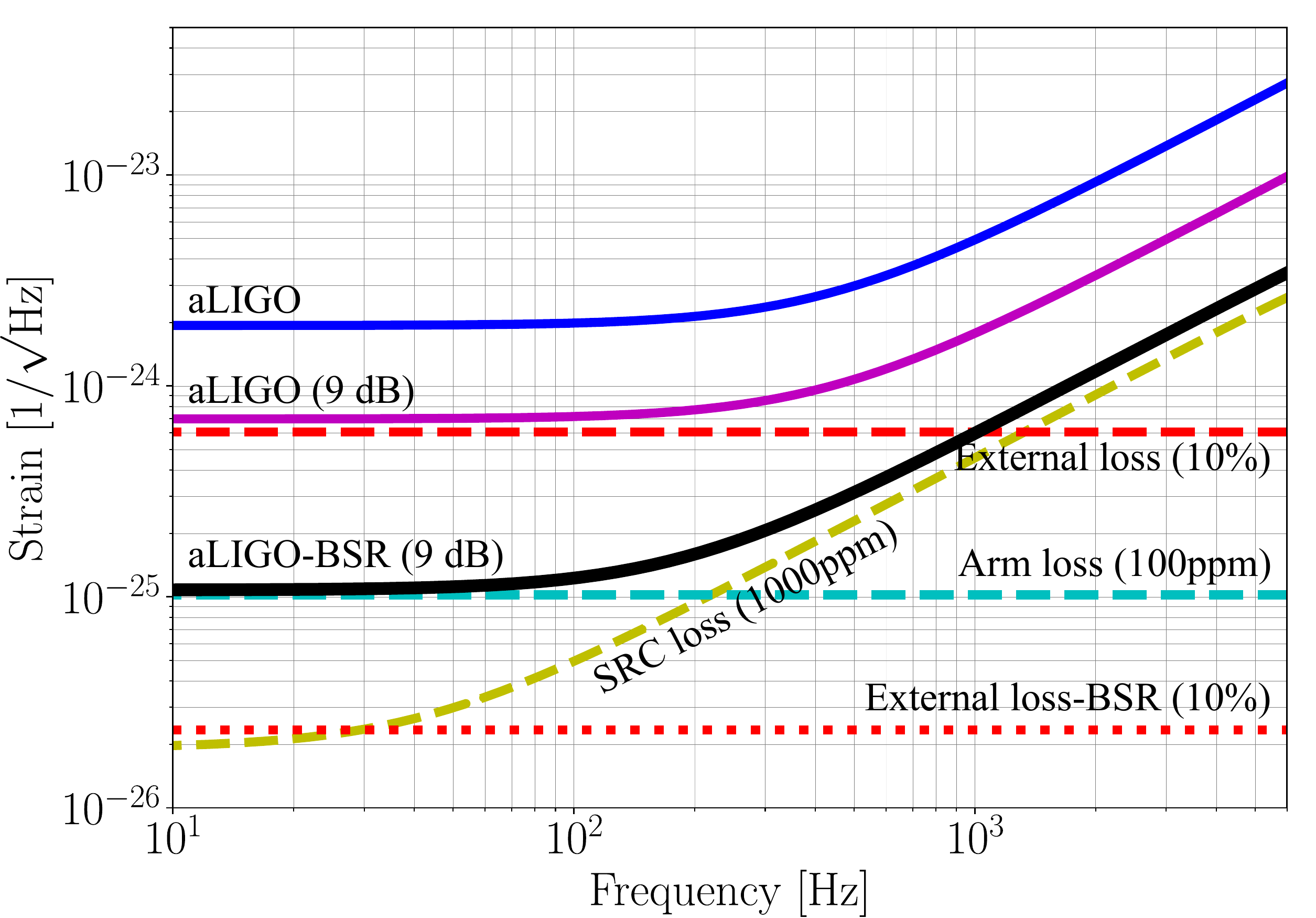}
\caption{This figure shows the noise spectral densities of optical losses and the shot noise limited sensitivity with parameters from aLIGO\,\cite{2015}. We assumed 1000\,ppm SRC loss, 100\,ppm arm cavity loss and 10\% external loss. The blue line illustrates the shot noise of aLIGO. The purple line is with $\sim$ 9\,dB observed squeezing (15dB squeezing in the arm cavities).  The black line illustrates the resulting shot noise of the aLIGO configuration with the BSR scheme discussed in this work. It approaches the SRC loss limit at high frequencies and arm cavity loss limit at low frequencies.}
\label{fig:losslimit}
\end{figure}

\begin{figure*}[t]
\centering
  \includegraphics[width=2\columnwidth]{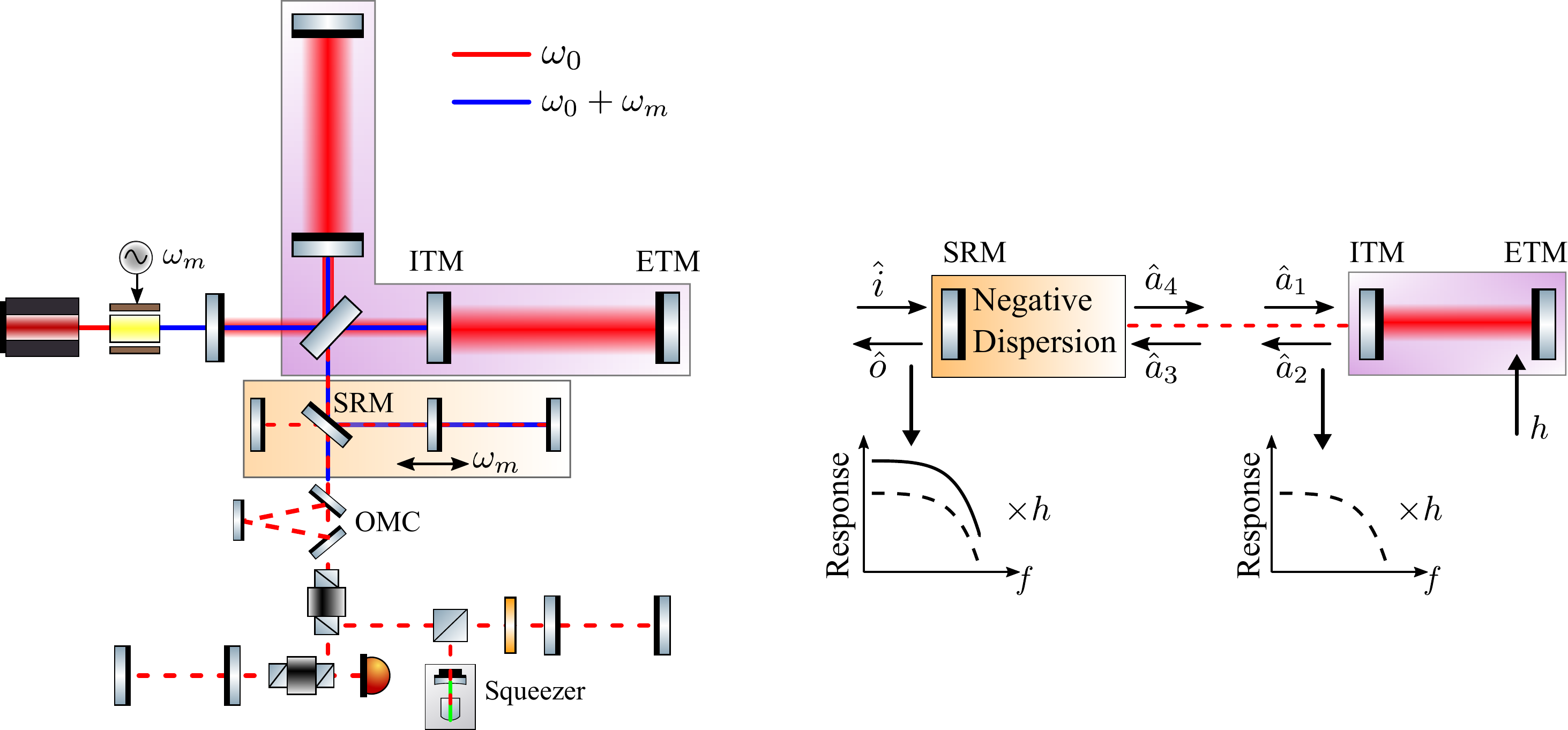}
\caption{Left: Gravitational-wave detector with the proposed BSR scheme and input/output filter cavities. The carrier laser is at $\omega_0$  (red). The optomechanical filter cavity is pumped with a laser at $\omega_0+\omega_m$ (blue), which comes from modulating the input carrier and transmits to the dark port under the Schnupp asymmetry. The resonant frequency of the mechanical oscillator is $\omega_m$. The blue beam is filtered with an output mode cleaner. The input and output filter cavities are used for frequency dependent squeezing and variational readout. Right: Simplified schematic of the broadband signal recycling scheme. Inside the SRC, the phase of sideband upon reflection from the arm cavity is cancelled, therefor the SRC is on resonance in a broad frequency band. The SRC is converted to be an amplifier for the signal coming out from the arm cavity with a gain of $1/\sqrt{T_{\rm SRM}}$.}
\label{fig:detector}
\end{figure*}
The optical losses are unavoidable in real detectors and can come from various sources,\,\textit{e.g.} absorption, scattering, mode mismatch and the open ports which are necessary to extract error signals for sensing and control. Depending on the mechanism how optical losses couple to the gravitational-wave channel, they can be classified into internal loss, including the arm cavity loss $\epsilon_{\rm arm}$ and signal-recycling cavity (SRC) loss $\epsilon_{\rm SRC}$, and external loss in the output chain $\epsilon_{\rm ext}$. The spectral density of these optical losses is given by\,\cite{PhysRevX.9.011053}: 
\begin{equation}\begin{split}\label{eq:qloss}
S_{\epsilon}=\frac{\hbar c^2}{4L^2\omega_0 P_{\rm arm}}\\
\left[\epsilon_{\rm arm}+\frac{(\Omega^2+\gamma_{\rm arm}^2) T_{\rm ITM }}{4\gamma_{\rm  arm}^2}\epsilon_{\rm SRC}+\frac{1}{4}T_{\rm SRC}\epsilon_{\rm ext}\right]\,,
\end{split}
\end{equation}
where $\omega_0$ is the laser frequency, $\Omega$ is the angular frequency of gravitational wave signals, $L$ is the arm length, $P_{\rm arm}$ is the arm cavity power, $T_{\rm ITM}$ is the input test mass (ITM) power transmissivity, $\gamma_{\rm arm}=cT_{\rm ITM}/(4 L)$ is the arm cavity bandwidth and $T_{\rm SRC}$ is the effective transmissivity of the signal-recycling cavity (SRC) formed
by ITM and the signal-recycling mirror (SRM).
The arm cavity loss limited sensitivity is frequency independent, since the vacuum couples into signals directly inside the arm. In contrast, the SRC loss limited sensitivity rises at higher frequencies above the arm cavity bandwidth due to the decrease of signal response. It is independent of any optical modules inside the SRC. The external loss happens on the output path of the interferometer, where there are unwanted effects including the mode mismatch at the output mode cleaner, and an imperfect quantum efficiency of the photodiode. The output loss noise depends on the SRC parameters and $T_{\rm SRC}$ equals to 0.14 in advanced LIGO (aLIGO). Apparently, it also limits the observed squeezing level. For example, with 10\% external loss, the observed squeezing cannot be more than 10\,dB. Fig.~\ref{fig:losslimit} shows the optical-loss limits of aLIGO\,\cite{2015}.

From Eq.~\eqref{eq:qloss}, we learn that at high frequencies, \textit{i.e.} $\Omega \gg \gamma_{\rm arm}$, the best achievable quantum sensitivity is limited by the SRC loss with spectral density,
\begin{equation}
S_{hf}=\frac{\hbar \Omega^2}{\omega_0P_{\rm arm}T_{\rm ITM}}\epsilon_{\rm SRC}\,,
\end{equation}
which turns out to be independent of the arm length. It is then exciting and crucial to explore the way of achieving the high frequency quantum limit, which will give the maximal astrophysical outcome, in particular for neutron star physics\,\cite{PhysRevD.98.044044,PhysRevD.99.102004}. As shown in Fig.~\ref{fig:losslimit}, the sensitivity of aLIGO with 9\,dB observed squeezing is still one order of magnitude worse than the SRC loss limit with $\epsilon_{\rm SRC}=1000\,$ppm in the frequency range of 1-5\,kHz.

Up to now, the proposed quantum techniques can be categorised into the following two classes: (1) noise suppression/cancellation schemes; (2) signal amplification schemes. The strategies that allow us to overcome the standard quantum limit are known as \textit{quantum nondemolition techniques}\,\cite{Braginsky547,yanbeiQND}, for example, frequency dependent squeezing and variational readout\,\cite{kimble2001}. Based on the understanding from the optical-loss limit, despite utilizing internal or external squeezing\,\cite{Adya_2020,PhysRevLett.118.143601,PhysRevLett.124.171101,PhysRevLett.124.171102}, the observed squeezing level has an unsurpassed bound due to external loss.  The signal enhancement scheme can potentially complement the squeezing for approaching the loss limit. The most well-known signal amplification technique, which has already been implemented in the second-generation detectors, is the SRC. Obviously, both the high-frequency and low-frequency sensitivities improve on \textit{resonant sideband extraction} mode. However, this comes with a price of sacrificing the peak sensitivity at intermediate frequencies. It is due to the accumulation of the phase of the sideband traveling inside the cavity (positive dispersion), which leads to a destructive interference when the frequency is larger than the cavity bandwidth. Another recent work on high-frequency detector design uses the coupled-cavity resonance of the arm cavity and the SRC. That scheme gives a narrow band dip in the noise spectrum at high frequencies\,\cite{PhysRevD.99.102004, ackley2020neutron}. 
Such a coupled-cavity resonance shows up when the signals resonate within the SRC. They can only resonate in a narrow high-frequency band. Equivalently, the identical resonance condition is satisfied around carrier frequency when the SRC is tuned on the \textit{signal recycling} mode.

One strategy to broaden up the bandwidth without sacrificing the peak sensitivity is to introduce negative dispersion to cancel the arm cavity round trip phase $2 \Omega L/c$, by means of white light cavities\,\cite{Rinkleff_2005,PhysRevLett.115.211104,page2020gravitational}. However, the negative phase provided by the unstable filter cavity in\,\cite{PhysRevLett.115.211104} evolves exponentially as a function of frequency.   As it turns out, there is an imperfect phase cancellation outside the frequency band of validity of the linear approximation to the exponential function, which eventually leads to rapid decrease of sensitivity above a certain frequency. 

The coupled-cavity and white light cavity ideas are heuristic. The scheme proposed in this work realises the coupled SRC and arm cavity resonance in a broad frequency band. We name it as the broadband signal recycling (BSR) scheme. 

\section{Design concept and quantum noise}

The scheme we propose in this work is shown in Fig.~\ref{fig:detector}. Inside the SRC, an optomechanical filter cavity which provides the negative dispersion is implemented for canceling the phase gained by the sideband upon reflection from the arm cavity. The optomechanical filter cavity is pumped with a laser at frequency $\omega_0+\omega_m$, which is generated from modulating the input carrier. A Schnupp asymmetry allows the carrier to transmit to the dark port. It is worthwhile to note the difference compared with the previous scheme in \cite{PhysRevLett.115.211104}. In that configuration, an addition mirror so-called iSRM is introduced for achieving an impedance match with the ITM, which effectively removes the ITM for the sideband. However, as we mentioned above, the negative phase provided by the optomechanical filter cavity cancels the round trip phase in the arms imperfectly at higher frequencies. Here we show the iSRM (extra complexity) is not required, which realises an exact phase cancellation in ideal case. 

\begin{figure}[h]
\centering
  \includegraphics[width=1\columnwidth]{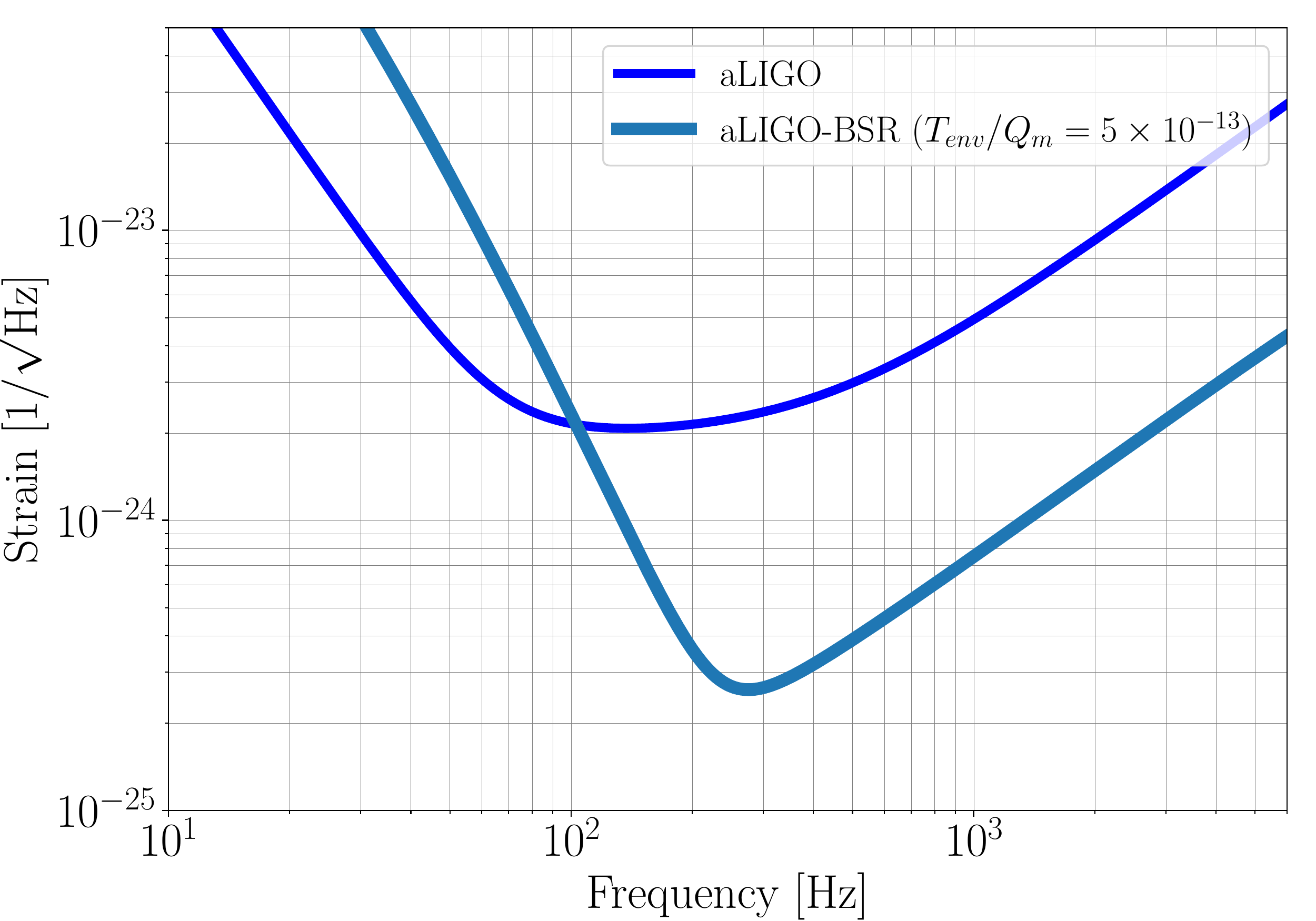}
\caption{This figure shows the quantum noise spectral densities of aLIGO and with BSR scheme ($T_{env}/Q_{m}=5\times 10^{-13}$). 15\,dB squeezing is assumed in the arm cavity and $\sim$ 9\,dB squeezing is observed under 10\% external loss.}.
\label{fig:RP}
\end{figure}

In Fig.~\ref{fig:detector}, the optical input/output relation of arm cavity can be derived as\,\cite{kimble2001}:
\begin{equation}
\hat{a}_{2}\approx\frac{\gamma_{ \rm arm}+i\Omega}{\gamma_{\rm arm}-i\Omega}\hat{a}_{1}=e^{i\beta}\hat{a}_{1}\,,
\end{equation}
where $\beta=2 {\rm atan}\frac{\Omega}{\gamma_{ \rm arm}}$ is the phase of the sideband when reflecting from the arm cavity. In order to cancel this frequency dependent phase inside the SRC and convert the SRC a broadband signal amplifier, we need to introduce negative phase lag.   
As it turns out, this can be exactly provided by an optomechanical filter in the so-called resolved-sideband regime, \textit{i.e.} $\omega_m\gg \gamma_f \gg \Omega$, where $\omega_m$ is the resonant frequency of the mechanical oscillator, $\gamma_f$ is the filter cavity bandwidth. When the mechanical damping rate $\gamma_{m}$, is much smaller than the negative mechanical damping rate due to the optomechanical interaction  $\gamma_{\rm opt}$, the optomechanical filter cavity gives the following open-loop transfer function\,\cite{PhysRevLett.115.211104}:
\begin{equation}\label{eq:optIO}
\hat{a}_{out}\approx-\frac{\gamma_{\rm opt}-i\Omega}{\gamma_{\rm opt}+i\Omega}\hat{a}_{in}=e^{i\alpha}\hat{a}_{in}\,,
\end{equation}
where $\gamma_{\rm opt}={P\omega_0}/({m\omega_m c L_{f}\gamma_{f}})$. There is $\gamma_{m}={\omega_m}/{Q_m}$. $m$ is the mass of the mechanical oscillator, $Q_m$ is the quality factor  and $P$ is the circulating power in the optomechanical filter cavity.  
When \begin{equation}
    \gamma_{\rm opt}=\gamma_{\rm arm}\,,
\end{equation} 
there is an exact cancellation 
of $\alpha$ and $\beta$ up to a frequency-independent constant:  $\alpha+\beta=\pi$. With the signal-recycling mirror tuned to give another constant $\pi$ phase shift, the short SRC is on resonance in a broad frequency band. As it turns out, $\hat{o}=\hat{a}_2/\sqrt{T_{\rm SRM}}$,
where $T_{\rm SRM}$ is the power transmissivity of the SRM.  
The effective transmissivity of the SRC is
\begin{equation}
T_{\rm SRC}= \frac{T_{\rm ITM}T_{\rm SRM}}{(1+\sqrt{R_{\rm ITM}R_{\rm SRM}})^2}\,,
\end{equation}
where $R_{\rm ITM}, R_{\rm SRM}$ is the power reflectivity of the ITM and SRM.  In the optomechanical filter cavity, the thermal noise contribution of the mechanical oscillator can be modelled as effective SRC loss\,\cite{PhysRevD.98.044044}:
\begin{equation}
\epsilon_{eff}=\frac{4k_B}{\hbar \gamma_{opt}}\frac{T_{env}}{Q_m}=1000\,{\rm ppm} \times \frac{0.014}{T_{\rm itm}}  \times \frac{L}{4\,{\rm km}} \times \frac{T_{env}/Q_{m}}{5\times10^{-13}\,K} \,,
\end{equation} 
where $k_B$ is the Boltzmann constant, $T_{env}$ is the environmental temperature. 
In our case, to guarantee an additional effective SRC loss contribution smaller than 1000\,ppm, $T_{env}/Q_{m}$ has to be  smaller than $  5\times 10^{-13}\,K$.
This gives an extremely  stringent requirement, which is the drawback of the BSR scheme compared with the scheme in\,\cite{PhysRevLett.115.211104}. This is because the  value of $\gamma_{\rm opt}$ in our scheme is smaller than that in \,\cite{PhysRevLett.115.211104}, where $\gamma_{\rm opt}=c/L$.  Therefore large bandwidth of the arm cavity is preferred. A $T_{env}/Q_{m}$ only $\sim 10^{-8}\,K$ is demonstrated in\,\cite{10.1038}. It is challenging to suppress the thermal noise in the BSR scheme.

After taking into account frequency dependent squeezing and variational readout as show in Fig.~\ref{fig:detector}, the spectral density of the detector can be calculated as the sum of\,\cite{Danilishin2019}
\begin{equation}\label{eq:S}
S=\frac{h_{\rm SQL}^2}{2}\left[\frac{e^{-2r}+\epsilon_{\rm ext}}{\mathcal{K}}+\frac{\epsilon_{\rm ext}}{1+\epsilon_{\rm ext} e^{2r}}\mathcal{K}\right]\,,
\end{equation}
and the SRC and arm cavity loss contribution, where $e^{-2r}\approx 0.03$ represents $\sim$ 15\,dB squeezing. 
$\mathcal{K}$ is the famous Kimble factor and $h_{\rm SQL}$ is the standard quantum limit,
\begin{equation}
\mathcal{K}=\frac{16 \omega_0P_{\rm arm}\gamma_{\rm arm }T_{\rm SRM}}{McL\Omega^2(\gamma_{\rm arm}^2+\Omega^2)}\,,\quad h_{\rm SQL}=\sqrt{\frac{8\hbar}{(M\Omega^2L^2)}}\,.
\end{equation}
Here $M$ is the mass of the detector test mass. 
We adopt $T_{\rm SRM}=0.005$ and $T_{\rm ITM}=0.014$, thus $T_{\rm SRC}=0.00002$. The arm cavity power $P_{\rm arm}$ is 800\,kW and arm length $L$ is 4\,km. The resulting shot noise limited sensitivity which corresponds to the sum of the first term in Eq.~\ref{eq:S} and the loss limit is shown in Fig.~\ref{fig:losslimit}. As we can observe, the shot noise limited sensitivity of the BSR scheme approaches the arm cavity loss limit at low frequencies and the SRC loss limit at high frequencies.  Including the radiation pressure noise and the oscillator thermal noise, the detector sensitivities are shown in Fig.~\ref{fig:RP}.

\section{Summary and outlook}

In this work, we show one scheme whose shot noise limited sensitivity can almost reach the optical-loss limit. An optomechanical filter cavity inside the SRC provides a proper negative phase lag that cancels the frequency dependent phase of the signal sideband upon reflection from the arm cavity. It allows the simultaneous resonance of broadband sidebands in the SRC, which can be viewed as a frequency dependent coupled-cavity resonance. Instead of suppressing the noise which will be limited by optical-loss, the signal is amplified in this scheme. The amplification gain is $1/\sqrt{T_{\rm SRM}}$. Note that the radiation pressure noise will also be amplified by the same amount, and the BSR scheme alone cannot overcome the standard quantum limit. For future detectors beyond the aLIGO, using high gain of BSR cavity provide a new way of reducing the shot noise, complementary to squeezing and increasing the power. A balance between the laser power and the signal recycling gain can provide easier access to cryogenic techniques for reducing mirror and suspension thermal noises, \textit{e.g.} cryogenic temperature which are proposed for LIGO Voyager\,\cite{Adhikari_2020}, the Einstein Telescope low-frequency detector\,\cite{ET-D,ET2020}, and Cosmic Explorer phase II\,\cite{Abbott_CE}. 

{\it Acknowledgements ---} 
T. Z., J.B. and H. M. acknowledge the support of the Institute for Gravitational Wave Astronomy at University of Birmingham. H. M. is supported by UK STFC Ernest Rutherford Fellowship (Grant No. ST/M005844/11).

\bibliography{bibliography}
\end{document}